\documentclass[preprint,aps,prd,amsmath,amssymb]{revtex4}
\usepackage{graphicx}

\begin{document}
\title{Statistical Hadronization and Holography}
\author{Jacopo Bechi}
\email[E-mail: ]{bechi@fi.infn.it}
\affiliation{CP$^{ \bf 3}$-Origins,
Campusvej 55, DK-5230 Odense M, Denmark.}
\begin{flushright}
{\it CP$^3$- Origins: 2009-27}
\end{flushright}
\begin{abstract}

In this paper we consider some issues about the statistical model of the hadronization in an holographic approach.
We introduce a Rindler like horizon in the bulk and we understand the string breaking as a tunneling event under this horizon. We calculate the hadron spectrum and we get a thermal, and so statistical, shape for it.

\end{abstract}

\maketitle

\section{Introduction}

A major phenomenon of the strong interaction is the hadronization. The hadronization is closely related to the confinement that is the less known aspect of the QCD. Quarks and gluons are not observable particles and in every physical process involving the strong interactions that is seen are colorless state: the hadrons. Because the hadronization is a intrinsically non-perturbative phenomenon, it is not calculable from first principles and one has to resort to phenomenological models.

A old model is the Statistical Hadronization Model (SHM). The idea of applying statistical concepts to the problem of the hadronization dates back to Fermi \cite{Fermi:1950jd} and further developed by Hagedorn \cite{Hagedorn:1965st}. More recently this approach obtained renovated attention in connection with the Quark-Gluon Plasma (QGP).

Basically the SHM assumes that the hadrons are produced from an excited region with probability depending only by the available phase space. In other words the hadrons occur like a completely equilibrated gas. Initially was shared the opinion that the hadrons thermalize in the QGP because multiple collisions; so in a kinematical way.

But the unexpected observation that also in elementary collisions the multiplicities of the observed hadrons have a thermal aspect \cite{Becattini:1995if}
 raised up a fundamental question. In fact, in elementary collisions, one can not invoke multiple rescattering, to explain the thermalization of the hadron gas because the secondaries are too few.

Eventually the SHM has been formulated in modern terms and applied to the issue of the hadron production in elementary collisions by a very successfull model (for a review \cite{Becattini:2009sc}). It is assumed that each high energy collision is followed by the production of hadronic clusters. A cluster is a colorless extended massive object labeled with abelian charge (electric, strange, etc.). The basic postulate of the model is that every multihadronic state localized within the cluster and compatible with the conservation law is equally likely. In other term every cluster can be described using a microcanonical ensemble and is assumed that, somehow, the hadrons be born in equilibrium; so the thermalization is stochastic.

As we already said the SHM is very successfull to reproduce the particle multiplicities but doesn't respond to the question how the thermalization is achieved.

Recently, an attempt to explain the universality of the thermal features in multihadron production has been put forward in \cite{Castorina:2007eb}
. This paper is formulated in term of a semiclassical picture of the hadrons like cromoelectric strings and is conjectured that the confinement can be represented as the QCD analog of a event horizon for colored signals. Then the string breaking following the pair production from vacuum is supposed similar to the Hawking-Unruh radiation from a Black Hole. In \cite{Castorina:2008gf}
there has been a tentative to understand the feature of this Black Hole. Postulating the hadron emission to be analog to a tunneling event trough a event horizon imply that the spectrum of the hadrons emitted is thermal because no information can be transfer with a tunneling between causally disconnected regions. The temperature of the radiation is related to the acceleration at the event horizon that is related to the string tension and so it is universal.

\section{Hadronization and Holography: motivations and set-up}

Because \cite{Castorina:2007eb}
 conjectures that the hadronization phenomenon is analog to a gravitational effect is natural to study the possibility to implement this idea in an holographic contest in which the QCD is described using a 5 dimensional dual gravitational theory (see also \cite{Evans:2007sf}
 for an application of statistical hadronization ideas to AdS/QCD).

The use of gravitational theories formulated in space-time with more than 4D to understand non perturbative aspects of gauge theories have begun with the pioneering Maldacena's work \cite{Maldacena:1997re}. In \cite{Maldacena:1997re} was conjectured that a superstring theory formulated in $AdS_{5}\times S_{5}$ is dual to Supersymmetric Yang-Mills $SU(N)$ theory with $\mathcal{N}=4$ supercharges formulated in the usual 4D Minkowski space-time. This theory is very different from QCD because is supersymmetric and conformal. Many efforts have been done to generalize the dual construction to more QCD like theories and the more successfull model including confinement and chiral symmetry breaking has proposed in \cite{Sakai:2004cn}
. However some important shortcoming remain to be solved.

At the same time another, more phenomenological, line of research started with the papers \cite{Da Rold:2005zs}\cite{Erlich:2005qh}
. In this approach, named bottom-up, one tries to build a 5D holographic description using the known features of the QCD and the holographic dictionary development for Conformal Field Theory.

A common limitation to both the approaches is that the gravitational side of the duality is tractable in case in which is valid the classical approximation. This correspond to a large $N$ limit in the gauge side, where $N$ is the number of colors.

In the early models \cite{Da Rold:2005zs}\cite{Erlich:2005qh}, in order to mimic the asymptotic freedom of the QCD, the 5D space-time has taken $AdS_{5}$

\begin{equation}\label{metric}
ds^{2}=\frac{R^{2}}{z^{2}}(-dt^{2}+d\mathbf{x}^{2}+dz^{2})
\end{equation}

where $R$ is the AdS radius and we put $R=1$ in the following. The confinement is implemented by hand putting a infrared  cutoff at $z=z_{IR}$. This implementation of the confinement is called Hard Wall (HW) and doesn't reproduce the linear Regge behavior of the hadronic resonance. The model has been improved in \cite{Karch:2006pv} replacing the sharp IR cutoff with a smooth dilaton profile. In this model, as expected from the classical description of hadrons like cromoelectric strings, the size of the particles increases with the mass. More recently a further step had done in \cite{Gursoy:2007cb}\cite{Gursoy:2007er}
that replace the by hand dilaton profile with a autoconsistent solution of the equations of motion for Einstein gravity coupled with the dilaton. In \cite{Gursoy:2007er} also has been shown that, requiring for confinement in the form of a area law for the Wilson Loop, one have to put a repulsive curvature singularity in the IR region.

In 4D the string breaking is a subleading phenomenon in $1/N$ expansion (in other words in the limit $N\rightarrow \infty$ the string cannot be broken). So, on a general ground, one have to expect that this effect cannot be described at the classical level in the gravitational dual. Of course the treatment of the full quantum effects in a gravity theory is over the our present possibilities. So one could wonder if the string breaking that occurs in the real world and that, following \cite{Castorina:2007eb}, is believed to be responsible for the thermal aspects of the hadronization can be studied in a holographic context. Indeed, as we will discuss later, one can argue that the hadrons production is naturally dual to a tunneling effect in AdS. A tunneling effect can be tackled using a semiclassical approach in first quantization without using the field theory derivation and without mention to loop effects in AdS. This tentative to translate the ideas of \cite{Castorina:2007eb} in holographic words is inspired by the description of the Hawking radiation using a semiclassical quantum mechanics tunneling picture that has been taken in \cite{Akhmedov:2006un}\cite{Akhmedov:2006pg}.

With these motivations in mind, now we seek to modify the holographic set-up to take in consideration the string breaking. Let's consider the HW model. Following \cite{Brodsky:2006uqa}
 the wall also can be understood as the maximum separation between the constituent quarks in a hadron. The approximation in this model is that the size of the hadrons is independent from the energy. In other words the cromoelectric string tension diverge when the separation $\xi$ between the quarks is $\xi=z_{IR}$.

But if the string is breakable, the hard wall have to be in some sense that can be overcome. So the HW have to be a description valid in the approximation $N\rightarrow\infty$. Moreover would be nice to replace the, not so well defined, wall with a proper gravitational, so metric, structure. Now, $1/N\sim\hbar$ (where $\hbar$ is the 5D Planck constant) because the large $N$ limit correspond to the semiclassical limit for colorless mesons, so the string breaking effect has to be described by a quantum effect in 5D. This argument motives us to replace the HW with a object that seems impenetrable at the classical level but not at the quantum level. The natural choice is a event horizon.

To be defined, we take the ansatz to replace the AdS metric \eqref{metric} with

\begin{equation}\label{metricadue}
ds^{2}=\frac{1}{z^{2}}\Big\{-\Big[1-az\Big(\theta(z-1/a+\epsilon)-\theta(z-1/a-\epsilon)\Big)\Big]^{2}dt^{2}+d\mathbf{x}^{2}+dz^{2}\Big\}
\end{equation}

where $a$ is a parameter to be fitted with the data and $\epsilon$ is a small parameter. This form of the metric is to understand only as a toy model and one could try to find a smother form. The important fact for us here is only the metric singularity that replaces the HW.

Beyond its simplicity, there is another reason to take the \eqref{metricadue} form of the metric. This metric is closely related to the Rindler metric. The Rindler metric is the space-time metric of a accelerating observer in Minkowski space-time. This metric has a horizon at $z=1/a$ and the radiation from this horizon is the cause of the Unruh temperature of the Rindler vacuum. From the 4D point of view of \cite{Castorina:2007eb} the string breaking is due to the fact that, after the on-shell production of the first vacuum pair (a QCD analog of the Schwinger mechanism of on-shell production of a $e^{+}e^{-}$ pair in a background electric field), Rindler horizon related to observer joint with one of the quarks of the pair cut the cromoelectric string causing the causal disconnection. Because this the string can break by the meaning of a tunneling trough a Rindler horizon. Then is natural to mimic the string breaking in 5D also by a tunneling through the Rindler horizon provided by \eqref{metricadue}.

It is worth to remark that, generally, the AdS/QCD models, differently respect the AdS/CFT, have to be understood as effective models builded to mimic some interesting feature of the QCD. Introducing an event horizon in the bulk we don't want to introduce a temperature on the 4D side. To be consistent, the 5D quantum field have to be considered to the classical level to avoid the problem of the conical singularity in the Euclidean Path Integral formulation of the theory.

\section{Hadronization and Holography: the calculation}

Let's consider now the holographic representation of the process:

\begin{equation}\nonumber
e^{+}e^{-}\rightarrow \gamma^{\star}\rightarrow q\bar{q}\rightarrow \text{hadrons}.
\end{equation}

Because the electron, the positron and the virtual photon are external degree of freedom respect QCD the first part of the reaction is localized at $z=z_{UV}$, where $z_{UV}$ is a ultraviolet regulator that can be taken arbitrary close to zero by a suitable renormalization procedure.

If we interpret, following \cite{Brodsky:2006uqa}, the $z$ coordinate as the separation between the primary quarks inside the hadron, the two parton can be holographically represented as point like particles that move in the direction of larger $z$. Of course, as they fly apart, the bare quarks radiate soft gluons and quarks/anti-quarks pairs. This process is known as showering and so the two point like particles in AdS represent dressed quarks. The important question for us is the probability $P(E)$ to find a dressed parton, with a energy $E$ at $z>1/a$. In fact, after the dressed quarks/anti-quarks pair have propagate at distance of order $1/\Lambda_{QCD}$, the confining dynamics of QCD become important and the quark pair forms a highly excited state that does colorless fragments. This fragmentation is called hadronization. So we interpret $P(E)$ as the probability that in the string breaking is product a meson state with energy $E$.

To calculate this probability we resolve the equivalent stationary problem in semiclassical approximation. For sake of simplicity we assume that the particles are scalar.

Let's consider the Klein-Gordon equation in the background \eqref{metricadue}:

\begin{equation}\label{eom}
\Big[-\frac{\hbar^{2}}{\sqrt{-g}}\partial_{M}g^{MN}\sqrt{-g}\partial_{N}+m^{2}\Big]\phi=0
\end{equation}

where $M,N$ indicate 5D indexes. The $m$ parameter is related to the mass of the parton.

We are looking for the solution of \eqref{eom} in the form $\phi(t,z)\propto\exp(-\frac{i}{\hbar}S(t,z))$, and at the first order in the limit $\hbar\rightarrow 0$. After a little of algebra one finds:

\begin{equation}\label{HJ}
-\frac{1}{(1-az)^{2}}(\partial_{t}S)^{2}+(\partial_{z}S)^{2}+\frac{m^{2}}{z^{2}}=0.
\end{equation}

Considering the stationary problem, we will look for solutions of \eqref{HJ} in the form $S=Et+S_{0}(z)$, where $E$ is the energy of the parton. Substituting in \eqref{HJ} we find

\begin{equation}
-\frac{E^{2}}{(1-az)^{2}}+\Big(\frac{dS_{0}}{dz}\Big)^{2}+\frac{m^{2}}{z^{2}}=0
\end{equation}

from which

\begin{equation}\label{differ}
\frac{dS_{0}}{dz}=\pm \frac{1}{(1-az)}\sqrt{E^{2}-\frac{m^{2}}{z^{2}}(1-az)^{2}}
\end{equation}

The equation \eqref{differ} has a singularity at $z=1/a$ on the real axis, so we have to define a contour of integration around the pole. From here we take, for simplicity, the chiral limit $m\rightarrow 0$; it doesn't afflict the our conclusions. The probability of the tunneling through the horizon $P(E)$ for the energy autostate with energy E is proportional to:

\begin{equation}
P(E)\propto \frac{|\phi(z_{B})|^{2}}{|\phi(z_{A})|^{2}}=\exp\Big[-\frac{2iE}{\hbar}\int_{C}\frac{dz}{1-az}\Big]
\end{equation}

where $z_{A}$ and $z_{B}$ are two point on the real axis respectively just before and just after $z=1/a$ and the contour $C$ is a half loop encircling the pole from below with extremities $z_{A}$ and $z_{B}$. With the change of variable $1-az=\rho$, $\rho=\epsilon e^{i\theta}$ and taking the limit $\epsilon\rightarrow 0$, we get

\begin{equation}\label{risultato}
P(E)\propto \exp\Big[-\frac{2E\pi}{a\hbar}\Big]
\end{equation}

This is a Boltzmann weight with a temperature equal to $T=\frac{a\hbar}{2\pi}$ to be interpreted like the universal hadronic temperature $T_{H}\sim 200 MeV$. This matching fixes $a$. According to the holographic dictionary the quantum effects in 5D are related to next order effects in 4D $1/N$ expansion and so the 4D interpretation of the 5D Planck constant is $\hbar\sim 1/N$. From \eqref{risultato} we can see, as expected, that in the limit $N\rightarrow\infty$ the tunneling probability tends to zero and so, in this limit, the string is unbreakable. Moreover, the hadronization temperature in naturally related to the confinement scale. After the breaking of the string, one could try to use the model of \cite{Evans:2009py} to study the following decay of the excited hadrons by meson radiation.

\section{Discussion and conclusions}

We conclude the paper noting the large N considerations in the holographic way, strongly support the idea of the hadronization as a tunneling process. Maybe this can be expected because, in some sense \cite{Yaffe:1981vf}
, the limit $N\rightarrow\infty$ is a classical limit. So one could suspect that subleading $1/N$ effects correspond to quantum effects. The use of the holographic correspondence makes this argument more precise.

\vspace{1cm}

\textbf{Acknowlegdements}: I would like thank the $CP^{3}-Origins$ for hospitality and the Fondazione Angelo della Riccia for financial support.

\end{document}